\documentclass[]{interact}

\usepackage{epstopdf}%
\usepackage{graphicx}
\usepackage[caption=false]{subfig}%
\usepackage{url}

\usepackage[numbers,sort&compress]{natbib}%
\bibpunct[, ]{[}{]}{,}{n}{,}{,}%
\makeatletter%
\def\NAT@def@citea{\def\@citea{\NAT@separator}}%
\makeatother%

\begin{document}

\articletype{Research Article}%

\title{Quantitative control and recording of materials-synthesis processes using an automated experimentation platform}

\author{
\name{Yusuke Hashimoto\textsuperscript{a}, Takaya Muramoto\textsuperscript{b}, Hikari Terada\textsuperscript{c}, Harim Song\textsuperscript{c}, Yuan Wang\textsuperscript{b,d} and Takaaki Tomai\textsuperscript{a,d}\thanks{CONTACT Takaaki Tomai. Email: takaaki.tomai.e6@tohoku.ac.jp}}
\affil{\textsuperscript{a}Frontier Research Institute for Interdisciplinary Sciences, Tohoku University, 6-3 Aramaki Aza Aoba, Aoba-ku, Sendai 980-8578, Japan; \textsuperscript{b}Graduate School of Engineering, Tohoku University, 6-6-11 Aramaki Aza Aoba, Aoba-ku, Sendai 980-8579, Japan; \textsuperscript{c}School of Engineering, Tohoku University, 6-6-04 Aramaki Aza Aoba, Aoba-ku, Sendai 980-8579, Japan; \textsuperscript{d}Institute of Multidisciplinary Research for Advanced Materials, Tohoku University, 2-1-1 Katahira, Aoba-ku, Sendai 980-8577, Japan}
}

\maketitle

\begin{abstract}
Data-driven materials development requires the collection of large amounts of high-quality materials data. Full autonomy of materials experiments is anticipated, but its technical hurdles are high and its adoption remains limited. In this study, we constructed a simple, easy-to-deploy automated experimentation platform that focuses not on full autonomy but on the reliable automation and quantitative recording of experimental processes. Specifically, commercially available instruments such as robot arms, electric pipettes, web cameras, and an electronic balance are combined, components such as fixtures are fabricated with a 3D printer, and the instruments are operated by control code generated by an AI agent based on a large language model. As a demonstration, we applied the platform to a two-solution mixing experimental system and synthesized ZIF-8, a metal--organic framework. A white suspension phase was observed in the product, and X-ray diffraction measurements confirmed that it was ZIF-8. We also found that its particle size distribution depends strongly on the solution dispensing speed of the electric pipette, which is a parameter that is difficult to control or record in manual operation. This dependence was reproduced in repeated runs, confirming the repeatability of the automated synthesis. This result is a good example showing that the control and recording of process parameters that are rarely quantified in manual work can govern the quality of materials data. All control code, CAD models, and documentation are made publicly available to encourage the spread of laboratory-scale automation of experiments.
\end{abstract}

\begin{keywords}
Laboratory automation; materials informatics; reproducibility; AI agent; process data; metal--organic framework; ZIF-8; open-source hardware
\end{keywords}

\medskip
\noindent\textbf{Impact Statement:} A simple open-source automation platform records and controls solution dispensing speed numerically, revealing a reproducible effect on ZIF-8 particle size that qualitative synthesis descriptions conceal.

\section{Introduction}

Realizing higher-performance devices requires the exploration of new materials and the enhancement of the performance and functionality of existing materials. Material properties are governed by composition, crystal structure, microstructure, synthesis processes, and other factors, and the number of their possible combinations is practically unlimited. Conventional materials development has relied on trial and error guided by the intuition, knowledge and experience of experts. This approach has therefore been enormously time-consuming and costly. To address this challenge, materials informatics (MI), which aims to dramatically accelerate materials development by combining data science with materials science, has attracted considerable attention in recent years~\cite{Rajan2005,Agrawal2016,Butler2018}. The Materials Genome Initiative (MGI), launched in the United States in 2011, promoted the development of materials-property databases and data-analysis infrastructure~\cite{MGI}. In particular, computational results such as first-principles calculation results have been collected and organized into materials databases such as the Materials Project~\cite{MP} and AFLOW~\cite{AFLOW}, which have served as a research and development foundation for MI and contributed greatly to its subsequent progress~\cite{Himanen2019}.

In device development, computational prediction of material properties alone is insufficient, and experimental validation is indispensable. Moreover, the effective use of data-driven methods requires a sufficient amount of experimental data. However, because materials experiments demand considerable time and labor, there is a strong demand for laboratory automation to acquire such data efficiently. In addition, data reproducibility is one of the major challenges in materials experiments~\cite{Baker2016}. Variability arising from different experimenters, apparatuses, and environments degrades the quality of experimental data and compromises the training and predictive performance of property-prediction models.

In response to these demands, research and development on autonomous experimentation systems combining machine learning and robotics has been actively pursued~\cite{Yachie2017,Roch2020,Stach2021,Seifrid2022,Ishizuki2023,Tamura2023}. For example, in 2020, Burger et al. reported an autonomous chemical experimentation system using a mobile robot, which performed 688 experiments over 8 days~\cite{Burger2020}. In 2023, GNoME, a deep-learning model developed by Google DeepMind, predicted hundreds of thousands of new stable crystal structures~\cite{Merchant2023}. Starting from these predictions, A-Lab combined machine-learning- and literature-based synthesis planning with robotics experimentation and succeeded in the automated synthesis of inorganic compounds in 17 days~\cite{Szymanski2023,Szymanski2026Correction}. These studies demonstrate that the integration of computational science, machine learning, and automated experimentation is a powerful approach to accelerating materials development.

On the other hand, most automated experimentation efforts in materials development pursue a high degree of autonomy, and the resulting systems have become increasingly complex. As a consequence, not only do initial costs rise, but specialized personnel with advanced skills in programming, CAD design, and robot control also become necessary, which constitutes a major barrier for laboratories and companies seeking to adopt automated experimentation~\cite{Kitchin2025}.

Against this background, we focus on reliable automation and process recording rather than full autonomy, and construct a simple, easy-to-deploy, open-source platform for automated experimentation. The platform lowers the programming barrier through the automatic generation of control code by an AI agent, and the CAD-design barrier by sharing a design bank of experimental components that can be fabricated with 3D printers. Furthermore, a monitoring mechanism based on a camera and IoT devices enables the quantification and automatic recording of experimental processes. All information required to reproduce the system, including the control code, CAD models, and related documentation, is made publicly available to encourage laboratory-scale automation of experiments. As a demonstration, we apply the platform to a two-solution mixing process for the synthesis of ZIF-8, a metal--organic framework, and aim to improve the reproducibility and efficiency of materials-synthesis experiments by achieving quantitative control of synthesis conditions and automatic recording of experimental process data. Automated ZIF-8 synthesis has been explored in complementary directions, including a remotely operated microfluidic synthesis~\cite{Wu2022} and automation-assisted synthesis combined with detailed recording of synthesis parameters~\cite{Smales2025}, and the dependence of ZIF-8 particle size on the precursor addition time has also been reported~\cite{Kim2024}. The present platform differs in that it automates conventional batch operations, such as pipetting, that are widely used in ordinary laboratories; using this platform, we control the addition speed numerically across the full setting range of the device and verify the repeatability of the resulting particle size distributions.

\section{Methods}

\subsection{Generation of control code for automated experiments by an AI agent}

\begin{figure}[htbp]
\centering
\includegraphics[width=\textwidth]{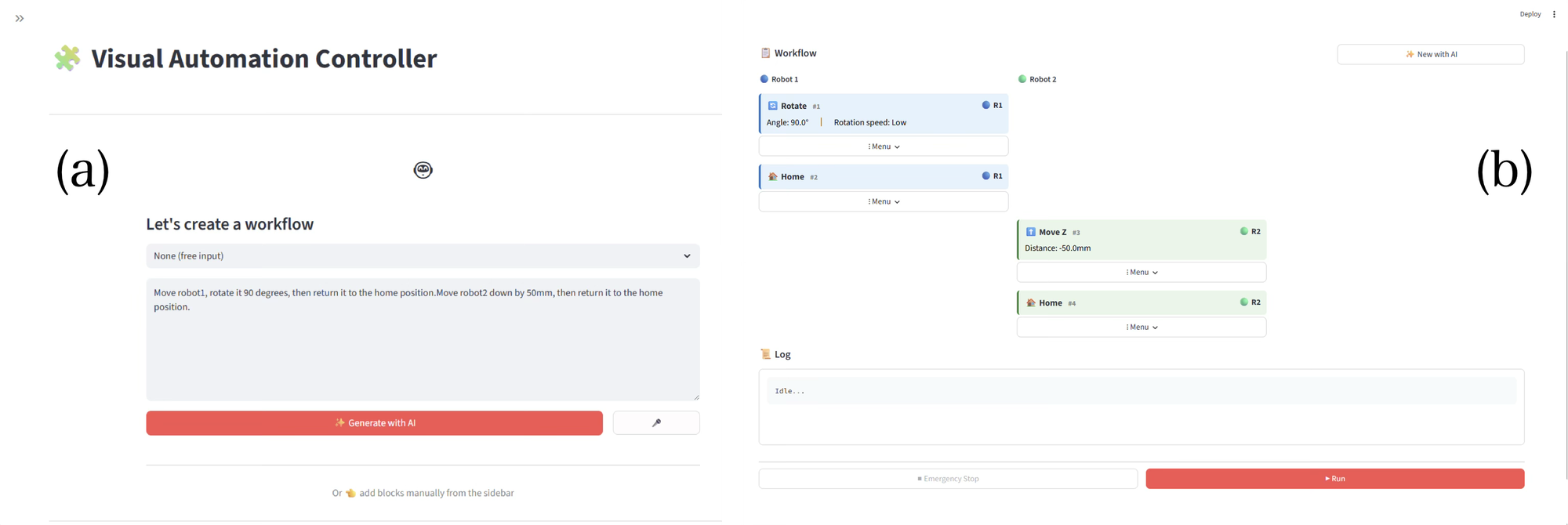}
\caption{(a) Automatic generation of control code by the AI agent. (b) GUI screen for confirming the experimental flow before execution.}
\label{fig:1}
\end{figure}

In materials experiments, the types, number, and arrangement of devices, as well as the process itself, are flexibly changed according to the purpose; in automated experiments, however, the control code must be adjusted every time the specifications change. We therefore developed a dedicated AI agent for efficiently generating control code on this platform, following recent demonstrations of large language model (LLM)-based agents in experimental sciences~\cite{Boiko2023,Yoshikawa2023,Febba2025,Xie2025}. The AI agent is built on Gemini 3.5 Flash (Google). Through prompts, it is provided with the functions, specifications, and communication protocols of each device constituting the automated experimentation system, together with the Python commands used to control them, and it automatically generates device control code according to the experimental objective (Figure~\ref{fig:1}(a)). The use of the AI agent enables flexible adaptation to changes in experimental equipment and environments. To improve the operational stability of the generated code and the reproducibility of experimental procedures, the AI agent outputs the experimental procedure as an experimental flow structured in JSON format. Constraining the output of the LLM to a structured JSON format suppresses variability in generation and enables programmatic validation before execution~\cite{Willard2023}. The output experimental flow includes the devices to be used, the order of operations, parameter settings, waiting times, and measurement conditions, and functions as an intermediate representation connecting the experimental objective given in natural language with the control code actually executed. The experimental flow is also visualized on a GUI, allowing the user to confirm its content before execution (Figure~\ref{fig:1}(b)). This makes it possible to verify the validity of the experimental procedure in advance and to reduce the risk of erroneous operations and inappropriate condition settings. Validated control code and experimental flows are registered as presets and can be reused according to the experimental purpose. Experimental parameters such as dispensing volume, temperature, stirring time, and measurement interval can be set via spreadsheet software, a browser-based GUI, or voice input.

\subsection{Setting experimental conditions by voice input}

\begin{figure}[htbp]
\centering
\includegraphics[width=0.65\textwidth]{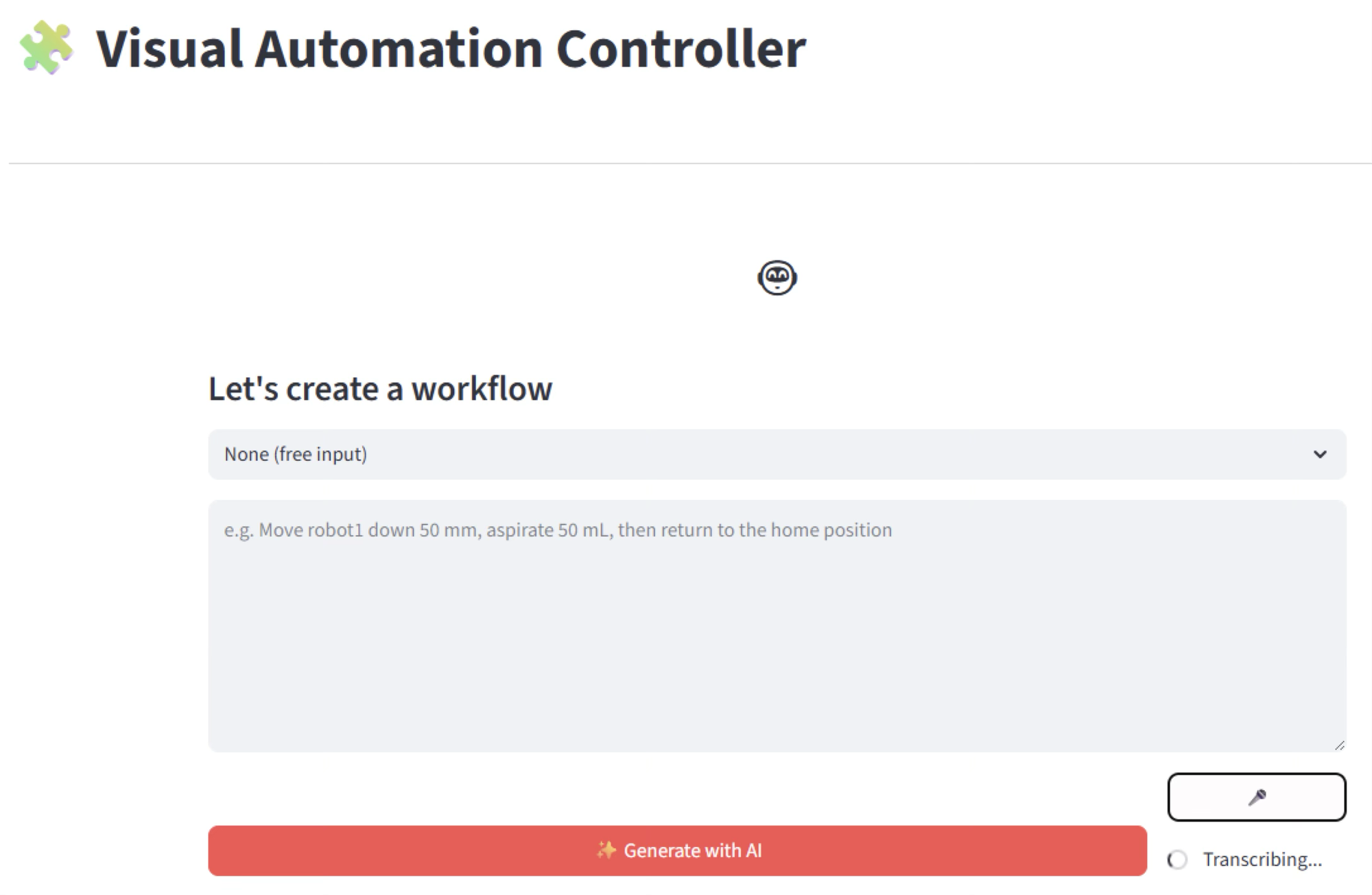}
\caption{Screenshot of the voice-input interface used to set an experimental flow.}
\label{fig:2}
\end{figure}

To enable changes to experimental settings when the operator's hands are occupied, the platform also supports voice input (Figure~\ref{fig:2}). Verbal instructions are converted into text by whisper-large-v3-turbo (OpenAI), a speech-recognition model, and the resulting text is fed into the AI agent described above and set as experimental parameters.

\subsection{Monitoring of experimental processes by a camera and IoT devices}

\begin{figure}[htbp]
\centering
\includegraphics[width=\textwidth]{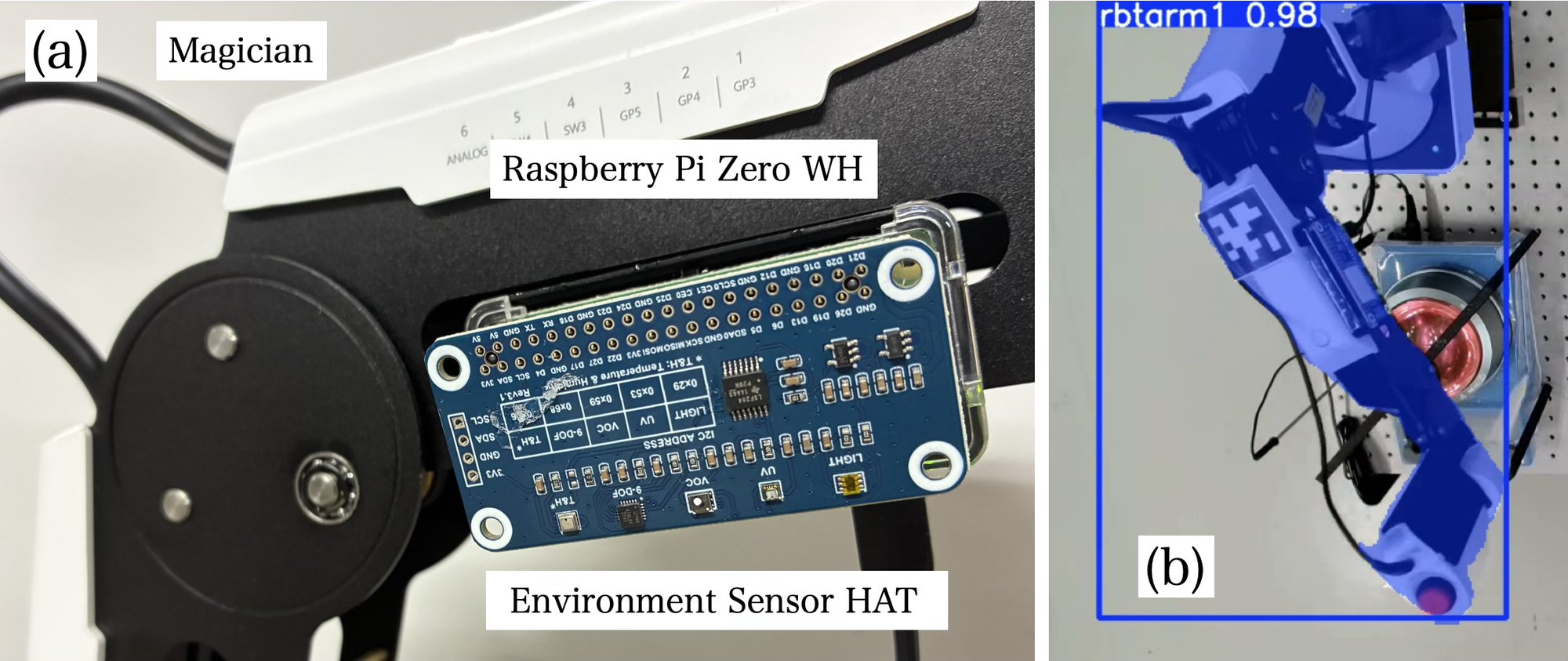}
\caption{(a) An IoT device consisting of a Raspberry Pi Zero WH and an Environment Sensor HAT mounted on the robot arm. (b) Camera image with YOLOv8 detection results overlaid.}
\label{fig:3}
\end{figure}

The processes of automated experiments are monitored with two IoT devices (one per robot arm) and a camera. Each IoT device consists of a Raspberry Pi Zero WH and a multi-sensor module (Environment Sensor HAT, Waveshare), and measures temperature, humidity, acceleration, angular velocity, and magnetic field (Figure~\ref{fig:3}(a)). The environmental and motion data obtained from these sensors are recorded as time-series data with timestamps, which facilitates their association with each experimental operation and subsequent analysis. In images acquired from the camera (C920n, Logitech) fixed above the system, the robot arms and electric pipettes are detected by YOLOv8~\cite{Terven2023} fine-tuned on images of these devices, and the state of the experiment is recorded as image data (Figure~\ref{fig:3}(b)). All process data, including the mass data obtained by the electronic balance, sensor data, and image-analysis results, are organized and saved as experimental logs.

\subsection{Design and sharing of experimental components via a CAD design bank}

\begin{figure}[htbp]
\centering
\includegraphics[width=\textwidth]{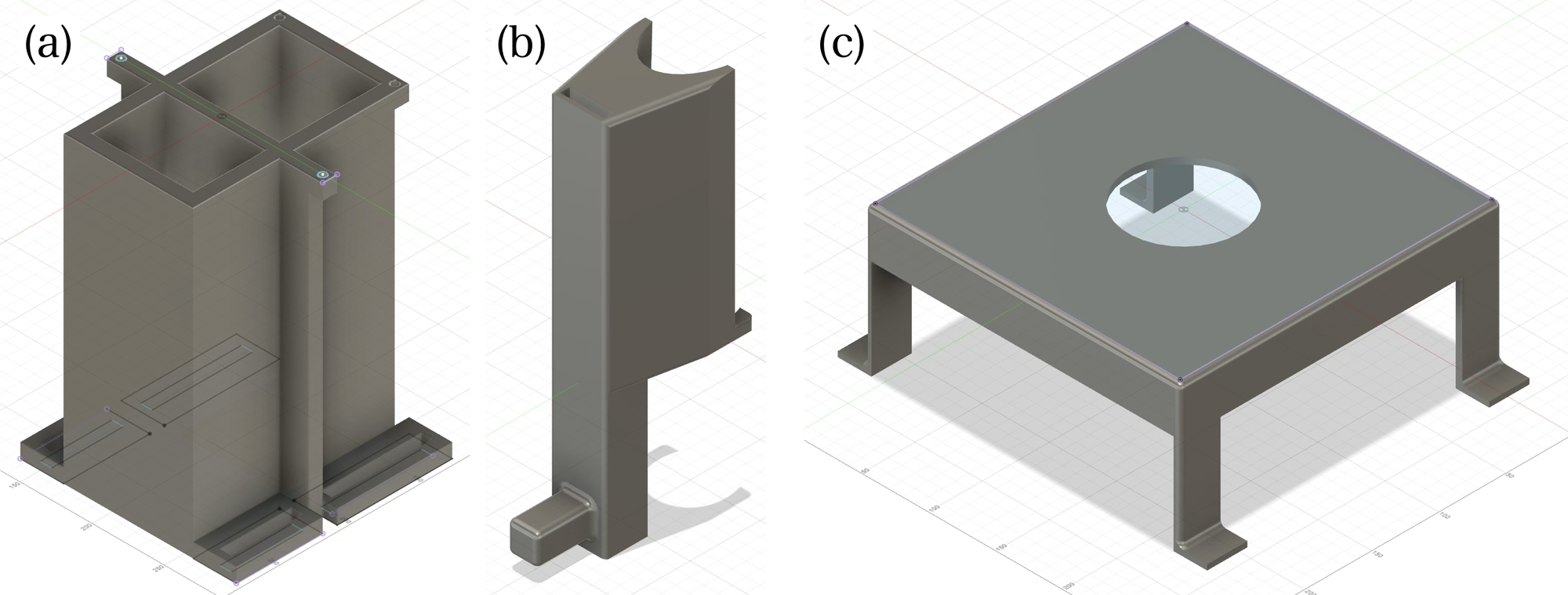}
\caption{3D CAD models of (a) the robot-arm holder, (b) the electric-pipette holder, and (c) the liquid-splash guard for the electronic balance (with windshield).}
\label{fig:4}
\end{figure}

The holders required to secure the flasks, electric pipettes, and robot arms, as well as a liquid-splash guard for the electronic balance (Figure~\ref{fig:4}), were designed with CAD software (Fusion 360, Autodesk) and fabricated with a 3D printer (X1E, Bambu Lab). This approach facilitates the modification of component shapes to suit each experimental setup and enables the low-cost construction of versatile and reproducible apparatus. The CAD models are organized as a design bank and released through a GitHub repository. Other researchers can use them to reproduce identical components and rapidly construct equivalent experimental setups.

\subsection{Configuration of the automated solution-mixing experimental system}

\begin{figure}[htbp]
\centering
\includegraphics[width=\textwidth]{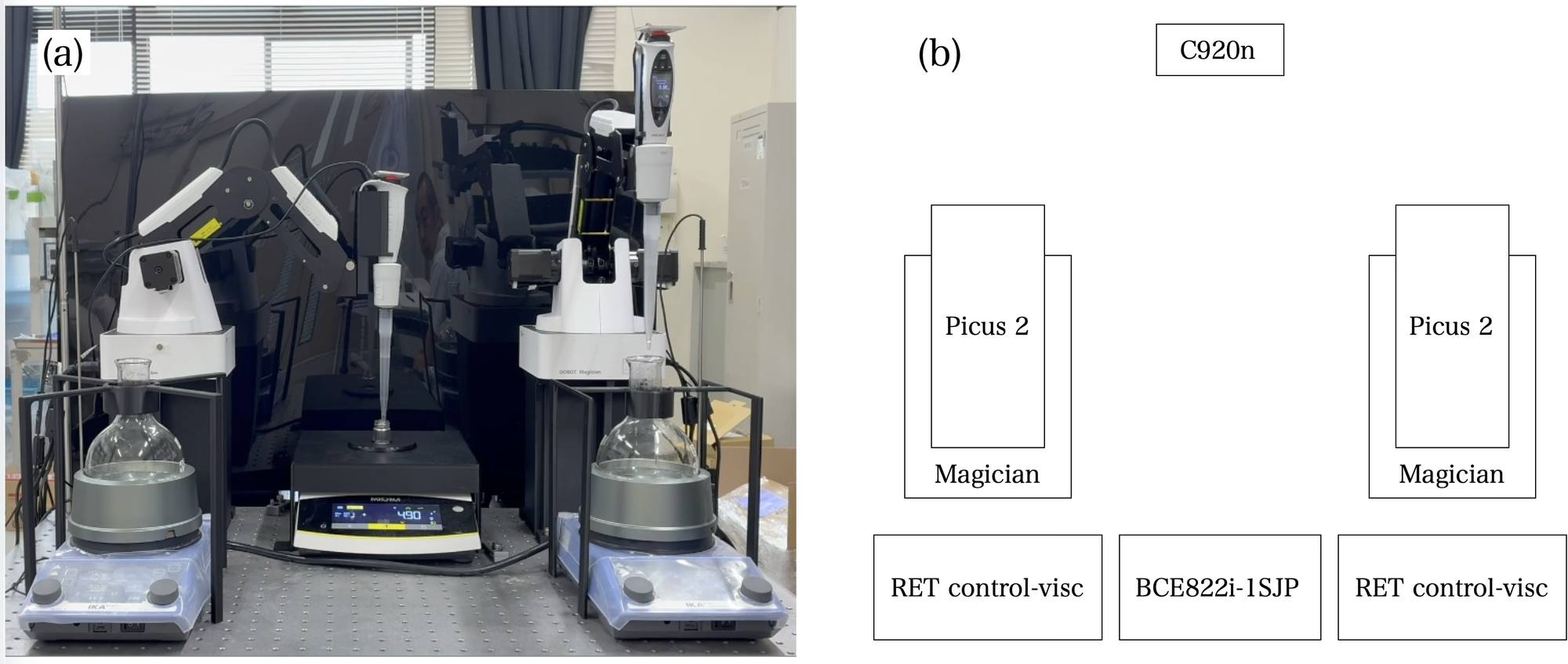}
\caption{(a) Photograph and (b) schematic of the automated solution-mixing experimental system.}
\label{fig:5}
\end{figure}

In this study, we constructed an automated experimentation system for dispensing and mixing two solutions (solutions 1 and 2). A photograph of the system is shown in Figure~\ref{fig:5}(a), and its schematic in Figure~\ref{fig:5}(b). The system mainly consists of two hot-plate stirrers (RET control-visc, IKA), two robot arms (Magician, Dobot), two electric pipettes (Picus 2, Sartorius; 500--10{,}000\,$\mu$L), and one electronic balance (BCE822i-1SJP, Sartorius; capacity 820\,g, readability 0.01\,g). In addition, the system is equipped with a camera and IoT devices for monitoring the experimental process. All devices are connected to a control PC via Wi-Fi or USB, and their operation and data acquisition are performed by Python programs.

Solutions 1 and 2 are prepared in flasks placed on the respective hot-plate stirrers, under stirring and temperature control. The temperature of the hot-plate stirrers can be set from room temperature to 400\,$^{\circ}$C. One robot arm and one electric pipette are assigned to each of solutions 1 and 2, which prevents cross-contamination between the solutions. Because each electric pipette handles only one solution, the tips do not need to be replaced. Each electric pipette is attached to the end of the corresponding robot arm, and its position is controlled by the arm. Each solution is aspirated with the electric pipette and dispensed into a receiving container placed on the electronic balance. The aspiration and dispensing speeds of the electric pipettes used in this study are electronically controlled in nine steps; according to the manufacturer's speed table, the operating time for the maximum volume of 10\,mL ranges from 10.2\,s (speed setting 1) to 0.9\,s (speed setting 9), corresponding to nominal average speeds of approximately 1--11\,mL/s~\cite{Sartorius2023}. The mass of the dispensed solution is measured by the electronic balance and recorded as the actual dispensed amount.

\subsection{Imaging system for appearance evaluation}

\begin{figure}[htbp]
\centering
\includegraphics[width=\textwidth]{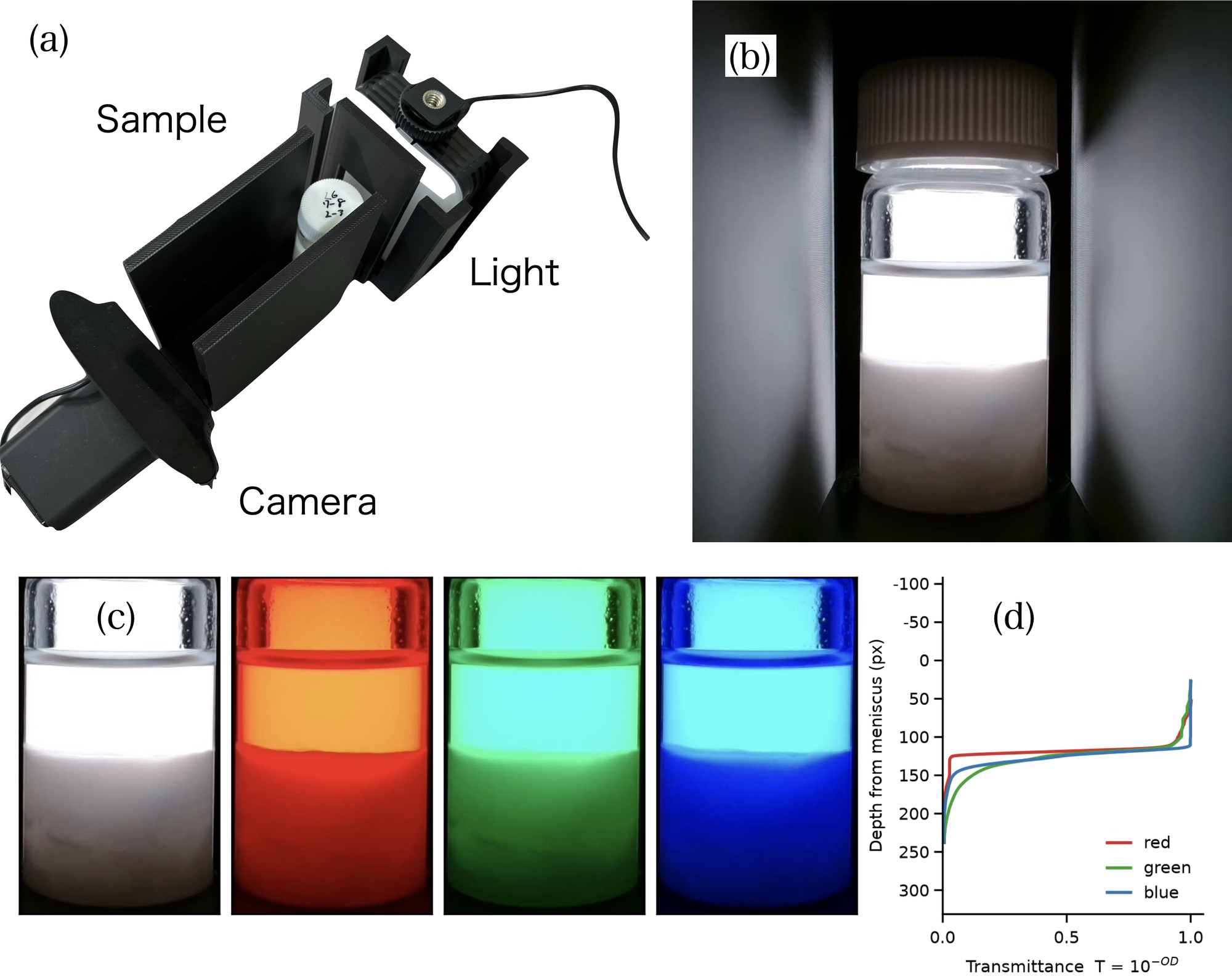}
\caption{(a) Photograph of the imaging system for appearance evaluation and (b) an image taken with the system. (c) Comparison of sample images taken under white, red, green, and blue illumination. (d) Depth profile of the transmitted-light intensity.}
\label{fig:6}
\end{figure}

In solution-mixing experiments, changes in appearance, such as suspension formation and discoloration of the products, also provide a great deal of information. In particular, such changes serve as a simple and rapid indicator for evaluating experimental reproducibility. In this study, we therefore used a purpose-designed system (Figure~\ref{fig:6}(a)) to acquire photographs under identical conditions for evaluation (Figure~\ref{fig:6}(b)). The system consists of a light (RGB62, NEEWER), a camera (C920n, Logitech), and a 3D-printed holding stand. The relative positions of the light, sample, and camera are fixed by the stand. Settings such as the intensity and color of the light and the focus position and exposure time of the camera are controlled by software so that the imaging conditions are kept identical across experiments. Photographs can be taken under white, red, green, and blue illumination, enabling a simple evaluation of the absorption characteristics of colored products (Figure~\ref{fig:6}(c)). Furthermore, arranging cropped sub-images of the same region of the photographs side by side (Figure~\ref{fig:6}(c)) and comparing transmitted-light intensity profiles along the depth direction (Figure~\ref{fig:6}(d)) enable intuitive and quantitative evaluation of, for example, interface positions and concentration fluctuations.

\section{Results}

\subsection{Automated synthesis of ZIF-8 (zeolitic imidazolate framework-8)}

\begin{figure}[htbp]
\centering
\includegraphics[width=\textwidth]{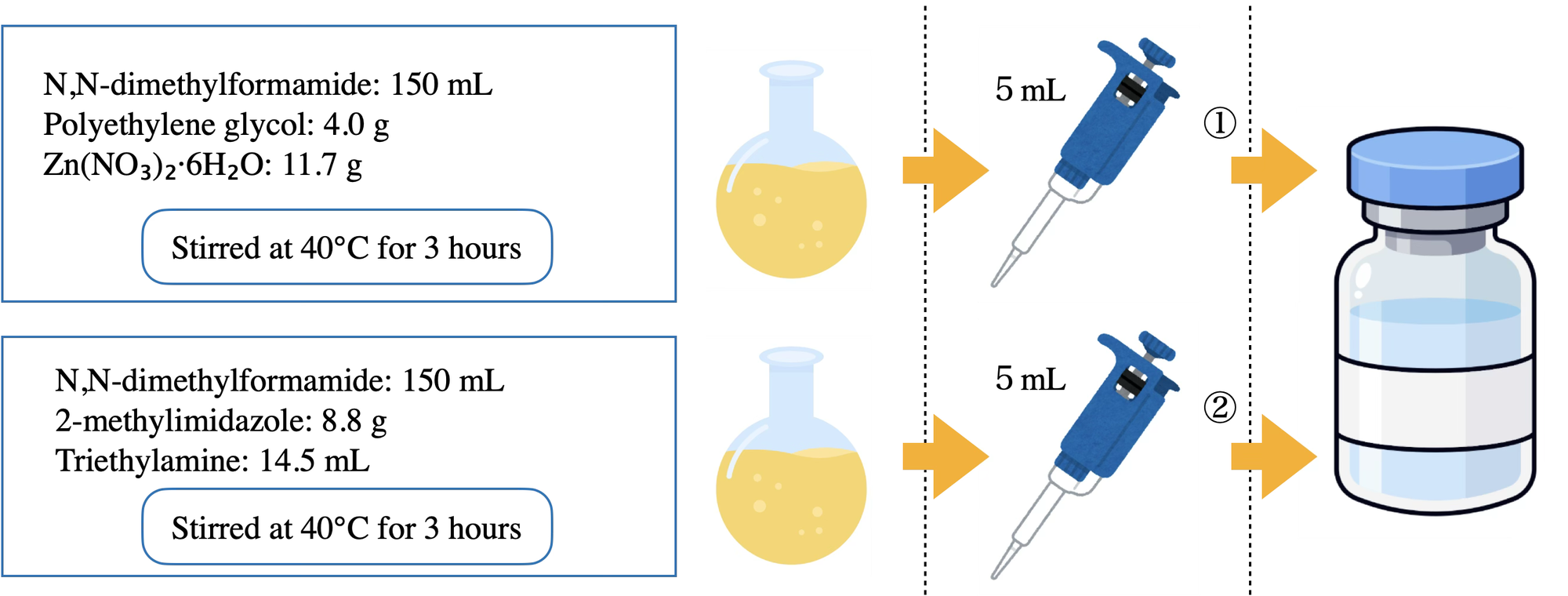}
\caption{Schematic of the synthesis procedure for ZIF-8.}
\label{fig:7}
\end{figure}

Using the present system, we synthesized ZIF-8~\cite{Park2006}, a metal--organic framework (MOF)~\cite{Furukawa2013}. MOFs have attracted attention as catalysts and gas-adsorption materials. The synthesis method was based on the report by Missaoui et al.~\cite{Missaoui2022}. Specifically, solutions 1 and 2, described below, were prepared and mixed in equal volumes (Figure~\ref{fig:7}). All reagents were purchased from FUJIFILM Wako Pure Chemical Corporation and used as received. Solution 1 was prepared by adding 4.0\,g of polyethylene glycol (PEG, average molecular weight 20,000; first grade) to 150\,mL of $N$,$N$-dimethylformamide (DMF; guaranteed reagent), followed by the addition of 11.7\,g of Zn(NO$_{3}$)$_{2}\cdot$6H$_{2}$O (guaranteed reagent). Solution 2 was prepared by dissolving 8.8\,g of 2-methylimidazole (2-mIm; first grade) in 150\,mL of DMF, followed by the addition of 14.5\,mL of triethylamine (TEA; guaranteed reagent). After both solutions were stirred at 40\,$^{\circ}$C for 3\,h, ZIF-8 was synthesized using the automated solution-mixing experimental system constructed in this study: 5\,mL of solution 1 was dispensed into a receiving container, and 5\,mL of solution 2 was then dispensed into the same container. After the injection of solution 2, the reaction mixture was allowed to stand at room temperature in capped vials for more than one week. The products were collected by centrifugation (10{,}000\,rpm, 4\,min), washed three times with DMF, and dried under vacuum at 60\,$^{\circ}$C for 12\,h. XRD patterns were recorded on a SmartLab X-ray diffractometer (Rigaku, Tokyo, Japan) equipped with a 9\,kW rotating-anode Cu~K$\alpha$ source, over $2\theta$ = 5--50$^{\circ}$. DLS measurements were performed on a Zetasizer Ultra Red (Malvern Panalytical, Malvern, UK) at 25\,$^{\circ}$C using DMF as the dispersant, and number-weighted size distributions were obtained by the instrument software.

In ten repeated experiments, the dispensed masses of solutions 1 and 2 were measured to be 4.80 $\pm$ 0.14\,g and 4.51 $\pm$ 0.12\,g, respectively, where $\pm$0.14 and $\pm$0.12 denote the standard deviations. The difference between the mean dispensed masses of solutions 1 and 2 arises from the difference in the density between the two solutions, which yields different masses even for equal dispensed volumes, as well as possibly from instrumental errors of the electric pipettes or from leakage caused by imperfect tip sealing.

A white suspension phase was observed in all synthesis experiments performed in this study. An image of ZIF-8 in a screw-top vial taken with the imaging system is shown in Figure~\ref{fig:6}(b). The white suspension phase of ZIF-8 was concentrated in the lower part of the vial, with a supernatant above it. This distribution was also clearly captured in the depth profile of the transmitted-light intensity (Figure~\ref{fig:6}(d)). In addition, because the present samples were white, no contrast difference due to the illumination color was observed (Figure~\ref{fig:6}(c)).

\subsection{Dispensing-speed dependence and repeatability of automated ZIF-8 synthesis}

\begin{figure}[htbp]
\centering
\includegraphics[width=\textwidth]{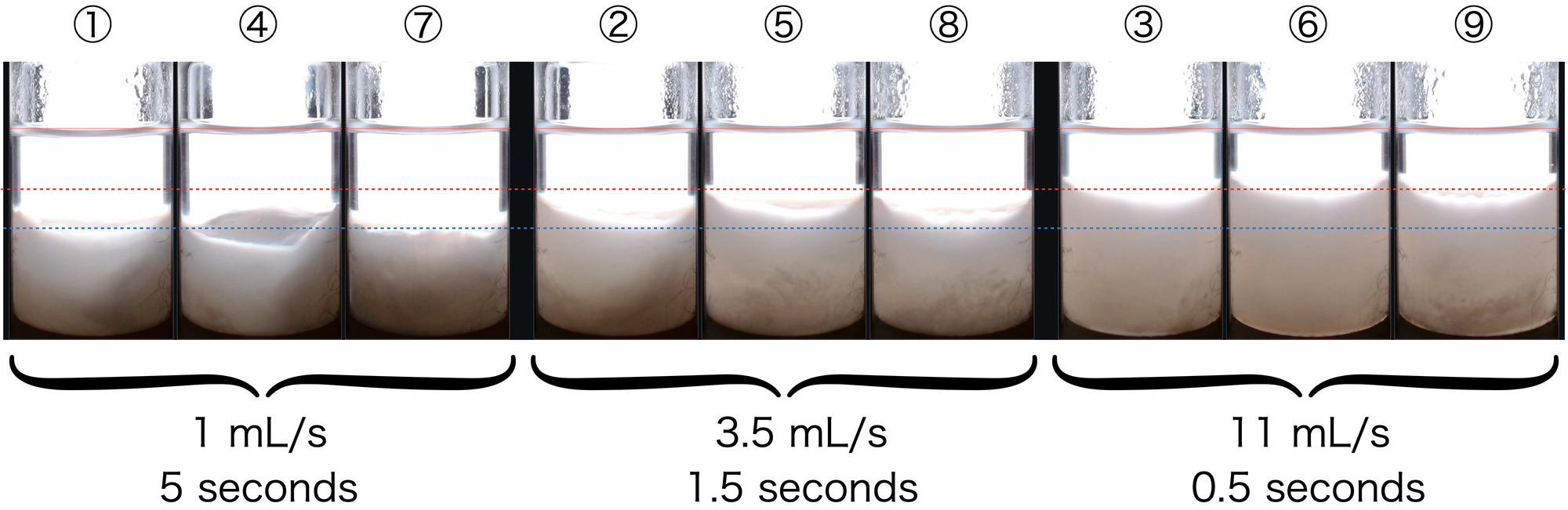}
\caption{Photographs of ZIF-8 suspensions in screw-top vials synthesized with different dispensing speeds of solution 2. The numeric labels indicate the order of preparation. The left, middle, and right columns correspond to speeds of 1, 3.5, and 11\,mL/s, respectively. The blue and red dashed lines are guides to the eye indicating the interface heights for speeds of 1 and 11\,mL/s, respectively.}
\label{fig:8}
\end{figure}

Electric pipettes allow the aspiration and dispensing speeds of solutions to be controlled consistently through numerical speed settings. We therefore set the dispensing speed of solution 2 to speed settings 1, 5, and 9 of the electric pipettes, which span its full setting range. According to the manufacturer's speed table, these settings correspond to operating times of 10.2, 2.9, and 0.9\,s for the maximum volume of 10\,mL~\cite{Sartorius2023}, that is, nominal average dispensing speeds of approximately 1, 3.5, and 11\,mL/s; for the 5\,mL dispensed in this study, the estimated dispensing times are approximately 5, 1.5, and 0.5\,s. The aspiration speeds of solutions 1 and 2 were fixed at speed setting 5, and the dispensing speed of solution 1 at speed setting 9. In addition, to verify repeatability, nine syntheses, three for each condition, were performed in an order that alternated among the three conditions. Photographs of the resulting suspensions are shown in Figure~\ref{fig:8}. All samples were photographed in a single session more than one week after the last synthesis. The numbers in the figure indicate the order of preparation, and the left, middle, and right columns correspond to dispensing speeds of 1, 3.5, and 11\,mL/s, respectively. The blue and red dashed lines indicate the interface heights for dispensing speeds of 1 and 11\,mL/s, respectively. Interestingly, the position of the interface between the suspension phase and the supernatant depended strongly on the dispensing speed. Despite the time separation between runs of the same condition imposed by the alternating order, the variation within each condition was much smaller than the differences between the conditions. This result suggests that the observed differences among the suspensions originate mainly from the dispensing speed of solution 2 rather than from random run-to-run variability or temporal changes in the solutions.

\begin{figure}[htbp]
\centering
\includegraphics[width=\textwidth]{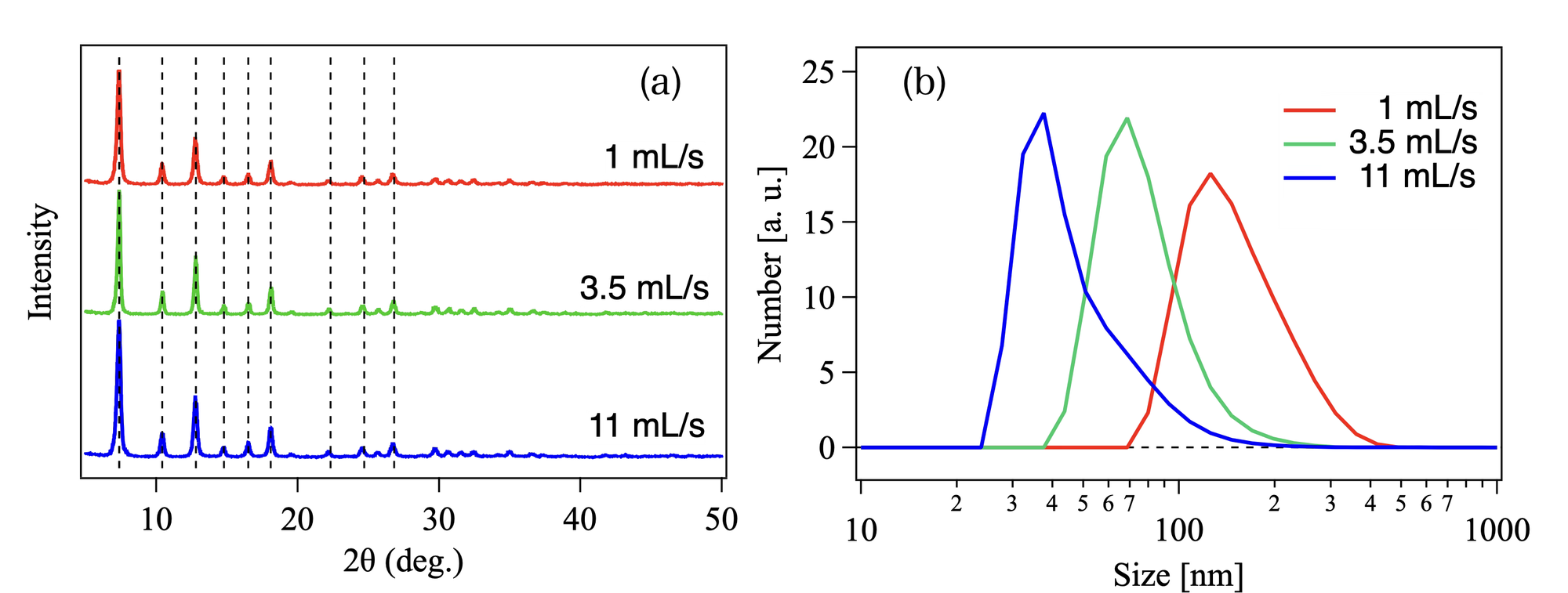}
\caption{(a) XRD patterns and (b) particle size distributions measured by DLS for the samples prepared with different dispensing speeds of solution 2.}
\label{fig:9}
\end{figure}

To quantitatively evaluate the effects of the dispensing time on the structure and particle size of ZIF-8, additional syntheses were performed for each condition; the washed and dried samples were subjected to X-ray diffraction (XRD) measurements (Figure~\ref{fig:9}(a)), and the samples redispersed in DMF were evaluated for particle size distribution by dynamic light scattering (DLS) (Figure~\ref{fig:9}(b)). The XRD patterns of all samples were consistent with ZIF-8 as the main crystalline phase under the tested conditions. The apparent Scherrer crystallite size, estimated from Gaussian fits to the (011) reflection ($K = 0.9$) without correction for instrumental broadening, was 25--36\,nm for all samples and showed no systematic dependence on the dispensing speed. In contrast, the particle size distributions measured by DLS showed a clear tendency for the ZIF-8 particle size to increase with increasing dispensing time, that is, with decreasing average dispensing speed. This tendency was reproduced in two independent syntheses for each condition and is consistent with the observations of the suspensions (Figure~\ref{fig:8}). These results demonstrate that the dispensing speed of solution 2 is an important parameter governing the particle formation of ZIF-8.

\section{Discussion}

The outcomes of materials-synthesis experiments depend on various experimental conditions, such as reagent concentration, temperature, stirring conditions, and addition speed. In the case of MOFs, the particle size is a practically important parameter: smaller particles are generally advantageous for catalysis and drug delivery because they offer larger accessible surface areas and shorter diffusion paths, whereas the particle size also governs interfacial compatibility and defect formation in membrane-based separation~\cite{Wang2026}. Precise control of the particle size during synthesis is therefore essential for optimizing the performance of MOFs across these applications. As described above, a pronounced dispensing-speed dependence was observed in the synthesis of ZIF-8; this tendency can be attributed to the effect of the dispensing speed on the balance between nucleation and crystal growth in the initial stage of the reaction~\cite{LaMer1950,Cravillon2011,Thanh2014}. In this scenario, under fast dispensing conditions, the precursors are mixed within a short time, so that the local concentration and supersaturation rise sharply and a large number of crystal nuclei are formed. Consequently, the precursors are distributed among many nuclei, the growth of individual particles is limited, and relatively small ZIF-8 particles are formed. Under slow dispensing conditions, in contrast, the precursors are supplied gradually and burst nucleation is suppressed; a larger fraction of the supplied precursors is consumed by the growth of existing nuclei, and larger particles are formed. We note that DLS probes the hydrodynamic size of particles redispersed after washing and drying, and therefore does not, by itself, distinguish primary-particle growth from changes in aggregation or dispersion state. Indeed, the apparent crystallite size was much smaller than the hydrodynamic size measured by DLS, suggesting that the ZIF-8 particles are polycrystalline assemblies of small crystallites and that the observed dependence of the particle size on the dispensing speed mainly reflects the assembly or aggregation of crystallites rather than the growth of individual crystallites. Thus, the observed size trend may reflect changes in nucleation, growth, assembly, and mixing during precursor addition, although the underlying mechanism was not directly resolved in this study. We also note that the opposite trend, smaller particles at longer addition times, has been reported for a surfactant-free aqueous synthesis with the reverse addition order~\cite{Kim2024}, and that both decreasing and increasing particle size with faster addition have been observed within a single automated study depending on the precursor concentration and addition order~\cite{Smales2025}. The effect of the addition speed on particle size therefore depends on the reaction system and cannot be generalized.

When pipettes are operated manually, the aspiration and dispensing speeds and the timing of addition tend to depend on the operator, leading to variability in experimental results. In the previous report on ZIF-8 synthesis referenced in this study~\cite{Missaoui2022}, the mixing operation is described as ``solution 2 was slowly added to solution 1,'' but no specific values are given for the addition speed or time. Such qualitative descriptions are one factor that makes experimental procedures difficult to reproduce. In contrast, in the present system, automating the solution dispensing operation with electric pipettes allows the aspiration and dispensing speeds and times to be set and recorded as numerical values. The three dispensing-speed settings used in this study span the full setting range of the electric pipette, and the dispensing-speed dependence and within-platform repeatability were evaluated at three conditions across this range. Note that a direct, quantitative comparison with manually performed synthesis was not conducted in this study; the evaluation presented here concerns the repeatability of the automated procedure itself. This result demonstrates the importance of controlling and recording, as numerical values, operating conditions that tend to be described only qualitatively in materials synthesis. It also illustrates that, for data-driven materials development, not only the structure of datasets but also the way materials experiments themselves are performed must be designed.

The challenge of reproducibility in materials experiments is not limited to a single laboratory~\cite{Baker2016}. It is unrealistic to always expect quantitative reproducibility among experimental data acquired by different experimenters, with different apparatuses, and in different environments. The automated experimentation platform presented in this study is simple yet highly versatile and reproducible, and can be deployed relatively easily. It can also record, in an integrated manner, liquid-addition conditions, temperature, stirring conditions, mass changes, sensor data, and image data. Furthermore, because all information necessary to reproduce the system, including the control code and CAD models, is provided as open source, many researchers can construct standardized experimental environments based on a common framework. If standardized experimental environments become widespread, mutually comparable experimental data can be accumulated across laboratories~\cite{Wilkinson2016}; this is a prerequisite for the systematic collection of experimental data into databases currently under way~\cite{Zakutayev2018,Katsura2019} and, by extension, for data-driven materials development. By enabling the collection of highly consistent experimental data, the platform is expected to serve as an experimental foundation for future materials development.

Based on the above, we summarize the key considerations for advancing laboratory automation as follows. First, no robot yet exists that possesses human-like flexibility in judgment and action and can therefore fully replace humans. Conversely, no human can perform pipetting operations---aspiration and dispensing---with exactly the same procedure and speed every time. It is therefore important to design systems in which humans and robots cooperate, compensating for each other's weaknesses while leveraging their respective strengths. Second, no matter how sophisticated a system is, it is meaningless unless it is actually used. It is essential to understand user needs through close communication and to build systems that continue to be used in the field. Third, the time available to system developers is limited. In particular, personnel who possess both knowledge of materials experiments and programming skills are scarce. It is therefore important to allocate limited resources on the basis of development difficulty and expected contribution to data quality, and to proceed with development systematically. For this reason, the present platform prioritizes reliable automation of selected parts of the experimental process over full automation of materials experiments.

Finally, we are extending this system to research on ``nanomaterials synthesis process informatics,'' which aims to efficiently explore the synthesis conditions of nanomaterials~\cite{Wang2026}. To date, we have developed an in-house database that links experimental conditions extracted from published papers with the properties of the resulting ZIF-8, and have constructed machine-learning models for property prediction. We are also experimentally verifying, with the present system, the ZIF-8 properties predicted for any given experimental conditions. The systematic accumulation of experimental operation data that are difficult to obtain from the literature alone, such as the dispensing speed, together with the resulting material properties, is expected to lead to machine-learning-based prediction of material properties and exploration of optimal synthesis conditions.

\section{Conclusions}

In this study, we constructed a simple, easy-to-deploy, open-source automated experimentation platform with the aim of improving the reproducibility and efficiency of materials-synthesis experiments. The platform consists of an AI agent that generates device control code, a mechanism for monitoring experimental processes with a camera and IoT devices, and a CAD design bank of experimental components. Using this platform, we performed the automated synthesis of ZIF-8, a member of the MOF family, and found that the particle size distribution of the resulting ZIF-8 depends strongly on the dispensing speed. These results demonstrate the importance of controlling and recording, as numerical values, operating conditions that have conventionally tended to be described only qualitatively, such as ``added slowly,'' and show that the platform is effective for the quantitative control and recording of materials-synthesis processes.

\section*{Data and code availability}

The program code, CAD models, and related documentation developed in this study have been made publicly available in a GitHub repository to facilitate their reuse and the reproduction of the platform (\url{https://github.com/YusukeHashimotoLab/saigen}).

\section*{Acknowledgements}

A part of this study was supported by Research equipment sharing system, Tohoku University (529, Multipurpose X-ray Diffraction Composite System (SmartLab 9MTP); 1079, Zetasizer Ultra (RED)). The authors used an AI assistant (Claude, Anthropic) to assist with language editing of the manuscript; all content was reviewed and verified by the authors.

\section*{Disclosure statement}

No potential conflict of interest was reported by the author(s).

\section*{Funding}

The author(s) reported there is no funding associated with the work featured in this article.

\section*{Declaration of generative AI use}

In accordance with the Taylor \& Francis policy on generative AI, the authors declare the following uses. Gemini 3.5 Flash (Google) was used to generate the device-control code as described in Section~2.1; whisper-large-v3-turbo (OpenAI) was used for speech recognition of voice commands. Claude (Anthropic) was used to assist with language editing of the manuscript. All AI-generated code and text were reviewed and verified by the authors, who take full responsibility for the content.

\section*{ORCID}

Yusuke Hashimoto \url{https://orcid.org/0000-0002-1886-9332}\\
Takaaki Tomai \url{https://orcid.org/0000-0003-0296-6565}

\section*{Author contributions}

Yusuke Hashimoto: conceptualization, methodology, supervision, writing -- original draft. Takaya Muramoto: investigation, software. Hikari Terada: investigation, software. Harim Song: investigation, software. Yuan Wang: investigation. Takaaki Tomai: supervision, funding acquisition, writing -- review \& editing.

\bibliographystyle{tfnlm}
\bibliography{refs}

\end{document}